\documentclass[twocolumn,showpacs,preprintnumbers,amsmath,amssymb,amsfonts,pre]{revtex4}

\usepackage{amsmath}
\usepackage{dcolumn} 
\usepackage{graphicx}
\usepackage{hyperref}
\usepackage{bm} 
\usepackage{subfigure}

\begin{document}

\title{Fame Emerges as a Result of Small Memory} 

\author{Haluk \surname{Bingol}}
\affiliation{Complex Systems Research Laboratory, Department of Computer Engineering, Bogazici University, Istanbul, Turkey}


\begin{abstract}
A dynamic memory model is proposed in which an agent ``learns'' a new agent by means of 
recommendation. The agents can also ``remember'' and ``forget''. The memory size is decreased while 
the population size is kept constant. ``Fame'' emerged as a few agents become very well known in 
expense of the majority being completely forgotten.  The minimum and the maximum of fame change 
linearly with the relative memory size. The network properties of the who-knows-who graph, which 
represents the state of the system, are investigated.
\end{abstract}

%

\maketitle

\section{Motivation}
One of the observations of complex systems is that they are made out of many interacting agents. 
Usually, the number of agents is simply too big for an agent to be informed of all the others. 
Therefore, agents act based on limited information. Many real-life examples can be given: A 
consumer can only have access to a limited number of suppliers. A car can only encounter a small 
number of other cars in a traffic jam. In the brain, a neuron cannot be connected to all the other 
$10^{11}$ neurons~\cite{Chialvo2004}. No web page can connect to all the other existing web pages. 
Similarly, no router can be connected to all other routers on the Internet. Even in many simple 
models, access to only the local information is a common property. In Bak's Sandpile Model, a sand 
particle communicates only with the four sand particles in its neighborhood~\cite{Bak1996}. 
Similarly, in Axelrod's 2D Culture Model, an agent interacts with its four neighbors 
only~\cite{Axelrod1997}. In Conway's Game of Life a cell checks its eight neighbors in order to 
decide whether to live or die in the next cycle~\cite{Berlekamp1982}. Although information exchange 
is relatively local and the rules of exchange are quite simple, these systems manage to become 
complex systems.

An individual cannot know the entire population but a small fraction of it. Consider the ratio of 
the number of people that one knows to the size of the population of the city or the country that 
she lives in. One expects that this ratio, which will be an important parameter of the model 
developed in this study, is a very small number~\footnote{For example, one may know $2 \times 10^3$ 
people while there are $15 \times 10^6$ in Istanbul that makes the ratio of $1.3 \times 10^{-4}$.}. 
Another observation is that the people that we know constantly changes. We ``learn'' new people 
from many sources including people, books, newspapers, radio, television, e-mail, www, SMS. 
On the other hand, we do not ``remember'' all the people that we learn. 
We have a limited cognitive capacity. 
A mechanism enables us to 
``forget'' people. Therefore, a model should deal with concepts such as population, memory, 
learning, remembering, forgetting and interaction of individuals that change their memory content. 
This paper mainly considers the human population in the development of the model but the findings 
are applicable to many systems. 

Mobile phones are a good example which satisfy many properties of the model presented in this paper
~\footnote{Example of mobile phone is proposed by one of the anonymous referee to whom the author would like to thank.}. 
They have a limited memory. 
When they receive a call, they try to store the caller number.
They usually do not store their own phone number, that is, they do not ``know'' themselves. 
Another example would be routers in computer networks.

\section{The Recommendation Model}
A reasonable question would be: What happens if an agent is allowed to interact with all the other 
agents, but remembers only a small fraction of them? In order to investigate this question a simple 
model is constructed. The dynamics of the system is investigated as the memory size is decreased.

\subsection{A Static Memory Model}
Let $\bm{a_i}$ be an agent. Let $\bm{A} = \{ a_i \: | \: 1 \leq i \leq n \}$ be a 
\textit{population} of $\bm{n}$ agents. Each agent $a_i$ has a memory $\bm{M_i}\subseteq A$. An 
agent $a_i$ \textit{knows} agent $a_j$ if $a_j \in M_i$.  The \textit{knownness} $\bm{k_i}$ of 
agent $a_i$ is the number of agents that know $a_i$. Then $k_i = \left| \{ a_j \: | \: a_i \in M_j 
\} \right|$. If everybody knows the agent, that is $k_i = n$, then the agent is called 
\textit{perfectly known}. On the other hand, if nobody knows the agent, that is $k_i = 0$, then the 
agent is called \textit{completely forgotten}. Knownness depends on $n$. For example, for $n=100$, 
one can be known by at most 100 people but for $n=1\,000$, knownness can be as high as $1\,000$. In 
order to compare populations of different sizes, a metric, independent of $n$, is needed. The 
\textit{fame} $\bm{f_i}$ of agent $a_i$ is defined as the ratio of its knownness to the population 
size, that is $f_i=k_i/n$. Since $0 \leq k_i \leq n$, it is always the case that $0 \leq f_i \leq 
1$. Hence fame is a normalized measure of knownness.

An agent \textit{learns} an agent $a_i$ if it gets $a_i$ in its memory. An agent $a_i$ 
\textit{remembers} agent $a_r$, if $a_i$ selects $a_r$ among the agents stored in its memory. An 
agent \textit{forgets} agent $a_f$ if it removes $a_f$ from its memory.

An abstraction which simplifies the model is made. It is assumed that every agent has the same 
\textit{memory size} $\bm{m}$, that is $ \forall i \, \left| M_i \right| = m$. Then the
\textit{total memory capacity} of the population is $n m$. The \textit{memory ratio} 
$\bm{\rho}$ is defined to be the ratio of memory size to the population size, $ \rho = m / n$. We 
have $0 \leq \rho \leq 1$, since this work considers $m$ values in the range  $0 \leq m \leq n$.  
Note that $\rho$ corresponds to the ratio of the number of people that one knows to the number of 
people one possibly know. As discussed in the ``motivation'' section, $\rho$ is expected to be a 
small value which corresponds to small memory sizes.

\begin{figure}
\sffamily
\rotatebox{90}{\begin{picture}(120,10)(0,0)\put(40,5){people}\end{picture}}
\begin{picture}(120,120)(0,0) \put(15,5){\line(1,0){80}} \put(15,110){\line(1,0){80}} 
\put(15,5){\line(0,1){105}} \put(95,5){\line(0,1){105}}

\put(40,125){memory} \put(5,100){1} \put(15,78){\line(1,0){80}} \put(5,70){g} 
\put(15,66){\line(1,0){80}} \put(15,43){\line(1,0){80}} \put(5,35){t} \put(15,31){\line(1,0){80}} 
\put(5,6){n}

\put(17,113){1} \put(40,113){r} \put(70,113){f} \put(88,113){m}

\put(50,66){\line(0,1){12}} \put(40,70){$a_r$} \put(38,66){\line(0,1){12}} 
\put(80,31){\line(0,1){12}} \put(70,35){$a_f$} \put(68,31){\line(0,1){12}}
\end{picture}

\caption{The state of the system is represented by an $n \times m$ matrix. 
The row $g$ corresponds to the memory $M_g$ of the agent $a_g$.
During recommendation, the $r^{th}$ item $a_r$ is selected by the giver agent $a_g$ and 
recommended to $a_t$. 
In order to make space for $a_r$, the taker agent $a_t$ selects the $f^{th}$ item $a_f$ to forget.
}
\label{fig:nmMatrix}
\end{figure}
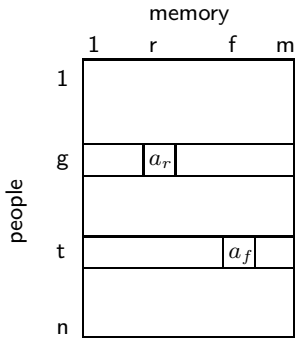

The \textit{state} of an agent is the content of its memory. Similarly the \textit{state} of the 
system is the memories of the all agents. The state of the system can be represented by an $n \times m$ matrix as in Fig.~\ref{fig:nmMatrix} where row $i$ corresponds to the memory $M_i$.

\subsection{A Dynamic Memory Model}
The system defined so far is a static one. In order to make it dynamic, interaction between agents 
is defined by means of recommendation. Agent $a_g$ \textit{recommends} agent $a_r$ to agent $a_t$ 
as visualized in Fig.~\ref{fig:nmMatrix}. The agents $\bm{a_g}$, $\bm{a_r}$ and $\bm{a_t}$ are 
called the \textit{giver}, the \textit{recommended} and  the \textit{taker}, respectively. The 
steps of recommendation process are i) $a_g$ remembers $a_r$; ii) $a_g$ gives $a_r$ to $a_t$; iii) 
$a_t$ learns $a_r$ if $a_t$ does not already know $a_r$. ``Remembering'' and ``forgetting'' are 
primitive operations. On the other hand, ``learning'' is not a primitive operation for $m<n$, since 
there is no empty space in the memory for the recommended agent. So learning comprises of three 
basic operations: i) remember some agent $a_f$ ii) forget $a_f$ in order to obtain an empty slot 
iii) put the recommended agent $a_r$ to this slot.

Some remarks about the recommendation operation is needed:

i) \textit{Selections.} The recommendation operation is carefully defined so that it is open to 
extensions. There are four selections in every recommendation operation, namely selections of 
giver-taker ($a_g$, $a_t$) and recommended-forgotten ($a_r$, $a_f$) as illustrated in 
Fig.~\ref{fig:nmMatrix}. These correspond to selections of $g$ and $t$ from the set of $\{1, 2, 
\cdots, n \}$ and memory positions $r$ and $f$ from $\{1, 2, \cdots, m \}$. Different 
specifications of selections would produce different results.  In the simple recommendation model 
of this paper, all four selections are defined to be randomly chosen from a uniform distribution.

ii) \textit{Axelrod's Culture Model.} Another selection criterion could be the case that both the 
giver and the taker should know the same people in order to interact. Restrict the selection of 
$a_g$ and $a_t$ in such a way that $| M_g \cap M_t | \geq k$, the case where two agents commonly 
know at least $k$ agents. For $k=1$, this leads to a model that is similar to Axelrod's Culture 
Model where culture vector has only one feature and the corresponding set of traits is $\{1, 2, 
\cdots, n \}$.

iii) \textit{Invariants.} The recommendation operation preserves some global values. Since there 
are $n m$ memory locations, the summation of the knownnesses of the system is given as 
$\sum_{i=1}^{n} k_i = n m$. This summation is invariant with respect to recommendation 
operation, since a recommendation increases the knownness of the recommended by one while decreases 
that of the forgotten by one.

iv) \textit{Completely forgotten.} If an agent becomes completely forgotten, then there is no way 
to be known again.

v) \textit{Perfectly known.} If an agent becomes perfectly known, it does not mean that it will 
stay this way unless the system is in one of its ``absorbing states''.

vi) \textit{Recommending items.} Note that although the memory model is presented as agents 
recommending agents, it can be extended to a model for a general case of agents recommending any 
type of items, such as books, songs, to other agents. Concepts such as ``completely forgotten'' 
would be difficult to explain for human population since a person would know herself even if the 
rest forgets her. On the other hand, it is not hard to talk about a song, a book, a cultural 
tradition or even a language that is completely forgotten. In science, there are many examples of 
concepts discovered, forgotten and re-discovered. A few changes would be needed to extend the 
model. Let $\bm{B}$ be the set of items. Then, an agent $a_i \in A$ would have items $b_j \in B$ in 
her memory, that is $M_i \subseteq B$. The memory ratio, that is the ratio of the actual number of 
memorized to the number of possibly memorized, would be $\rho = m / |B|$. The recommendation 
operation would be defined as an agent $a_g$ recommends item $b_r$ to agent $a_t$. In the rest of 
the paper, we assume that agents recommend agents to agents, that is $B = A$.

\subsection{A Simple Recommendation Model}
Many models can be built on these concepts. One of the simplest models, called the \textit{Simple 
Recommendation Model}, is obtained by defining all the selection mechanisms as random selections. 
There are four random selections for each recommendation. The giver, $g$, and the taker, $t$,  are 
selected randomly from the set of $\{1, 2, \cdots, n \}$. The giver $a_g$ selects the recommended 
agent $a_r$ from its memory by selecting $r$ from $\{1, 2, \cdots, m \}$ randomly. This is the 
remembering process. If the taker $a_t$ already knows $a_r$, then it does nothing. Otherwise it has 
to learn it. Learning calls for selecting a memory slot. This selection of $f$ is also done 
randomly from $\{1, 2, \cdots, m \}$.

This definition implies a number of properties. i) The selection rules do not prefer one agent to 
another. That is, the process is symmetric with respect to agents. ii) Any agent can get a 
recommendation from any other agent. Note that this may be an over simplification, since in real 
life examples an agent can get in touch with only a limited number of agents. On the other hand, 
increase in communication (e.g. via e-mail) may enable one to communicate with almost anybody.

\subsection{Termination Conditions}
When to terminate a simulation is a difficult issue. Defined this way, the memories of the agents 
are kept changing as long as the recommendations continues. There are some special cases in which 
continuing recommendations cannot change the state of the system. In these cases, the simulation 
can be terminated.

\textbf{Absorbing States.}
A state where every agent has exactly the same memory content, that is $\forall i,j \:  M_i = M_j$, 
is called an \textit{absorbing state}. In an absorbing state case, nobody can recommend anything 
new since everybody knows exactly the same $m$ agents and the remaining $n-m$ agents are completely 
forgotten. So there is no point continuing simulation. Therefore an absorbing state is a 
termination point. Since the system asymptotically converges into one of these absorbing states, 
absorbing states are theoretical terminations points. Note that $m=0$ and $n = m$ cases are special 
cases of absorbing states.

Simulations show that there are two regimes in the system~\cite{Bingol2005}. In the beginning, the 
system tends to forget. This forgetting mechanism works so powerful that many agents become 
completely forgotten at the very early stages of the simulation. As simulations proceeds, the 
number of known people becomes much less than the population size. Then system reverses this 
behavior. This time it tries not to forget. This is an expected behavior since the system converges 
to an absorbing state asymptotically. In this paper the second regime is investigated. The 
\textit{average recommendation per agent} $\bm{\nu}$ is defined to be the ratio of total number of 
recommendations to the population size $n$. Throughout this study $\nu=10^6$ is used.

\section{Random Initial Memory}
The initial configuration of the agent memory is important for the model. The memories of the 
agents are initially filled with randomly selected indexes from $\{1, 2, \cdots, n \}$. Repetitions 
are not allowed. The model is simulated for different values of $n$ and $m$. Population sizes of $n 
= 10^2, 10^3, 10^4$ and memory ratios of $\rho = 0.50, 0.30, 0.20, 0.10, 0.05, 0.01$ are used.

\begin{figure}
\includegraphics[width=\columnwidth,height=\columnwidth,keepaspectratio=true]{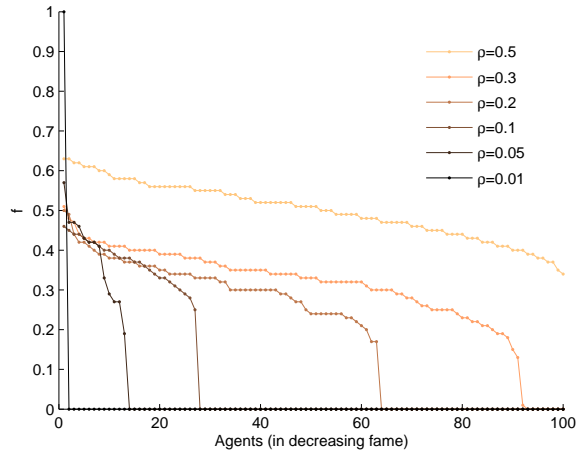}
\caption{(Color online) Even distribution of fame degrades as $\rho$ decreases. 
The model is simulated for various $m$ values where $n=100$ and $\nu=10^6$.
At the end of the simulation, memory dumps of agents provide who-knows-who information.
Fame of each agent is calculated and for better visualization, the agents are sorted in decreasing order in fame.
}
\label{fig:fig_effectOfRho}
\end{figure}

\subsection{Effects of memory size}
For the same population size $n$, effects of changing memory size $m$ in the interval $0 < m < n$ 
is investigated. Since the memories of the agents are initially randomly filled, the initial fame 
of an agent is around the average value of $<\!\!f\!\!>\,= m/n$. As $m$ is decreased, some agents 
become more known than others, at the expense of others becoming less known. Further decrease of 
$m$ increases the degeneration further.

For $n=100$, $\rho$ is changed and the change in fame $f$ is observed. 
Fig.~\ref{fig:fig_effectOfRho} is an example of various simulations which produce similar results. 
In this visualization, the agent number $1$ is the most famous one and the agent number $100$ the 
least famous. Note that the area under the curve is equal to the total memory capacity $n \times 
m$. As $m$ decreases, agents on the right become completely forgotten, as a result the agents on 
the left become increasingly famous. Around $\rho = 0.5$, the knownness of some agents becomes very 
low. Completely forgotten agents starts to appear around $\rho = 0.35$. From that point on, 
decrease in $m$ increases the number of completely forgotten agents. Since the total memory 
capacity is fixed, a few agents become very well known as a result of this process. Hence, fame 
emerges as an effect of small memory size.

Eventually $m$ decreases to the extreme case of $m = 1$ where an agent can remember only one agent. 
In this $\rho=0.01$ case, the dynamics of the system goes to an extreme. All the agents become 
completely forgotten, except only one. That lucky agent is known by the all other agents. This is 
the expected absorbing state since the number of known agents is $m=1$. In order to check this 
finding, simulations with larger values of $n$ is done for $m = 1$. It is observed that as the 
population size gets larger; reaching an absorbing state becomes harder.

\section{Regular Initial Memory}
One may suspect that these findings are due to small fluctuations of the random initial memory. 
Although random initialization does not prefer any agent systematically, it has some statistical 
variation. As a result of that, some agents could be slightly more known then others. This initial 
unbalance may affect the dynamics. In order to check this possibility, a perfectly symmetrical 
memory initialization scheme is used. In the regular initial memory scheme, each agent $a_i$ is 
allowed to know its $m$-neighbor, that is $M_i = \{ a_k \: | \: k \equiv i+j \pmod{n}$ for $1 \leq 
j \leq m  \}$, similar to the case of~\cite{Watts1998}. In this way, it is guaranteed that the 
knownness of every agent is exactly $m$.

For regular initial memory, $n=100, \cdots, 1\,000$ range with 100 increments is simulated. For 
each $n$, $\rho=0.10, \cdots, 0.90$ range with 0.05 increments is studied. Additionally, 
$\rho=0.01, \cdots,0.05$ range with 0.01 increments are simulated in order to see the behavior at 
very small values of $\rho$. Every $n$ and $\rho$ combination is simulated 20 times. Interestingly, 
both random and regular initial memory strategies produce similar results.

\begin{figure}[htbp]
\centering
\subfigure[] 
{
    \includegraphics[width=\columnwidth,height=\columnwidth,keepaspectratio=true]{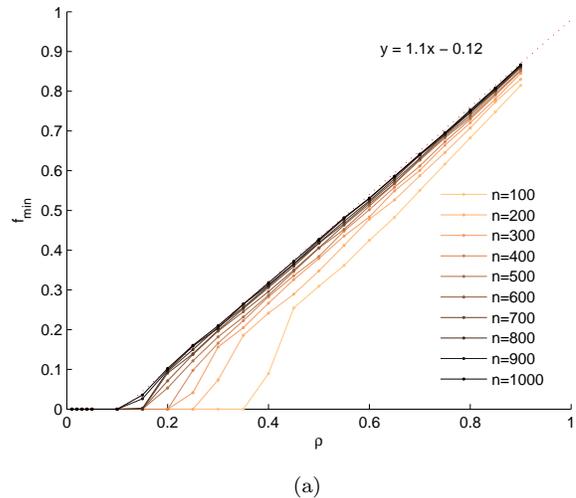}
    \label{fig:fig_rho_fminA}
}
\vspace{.1in}
\subfigure[] 
{
    \includegraphics[width=\columnwidth,height=\columnwidth,keepaspectratio=true]{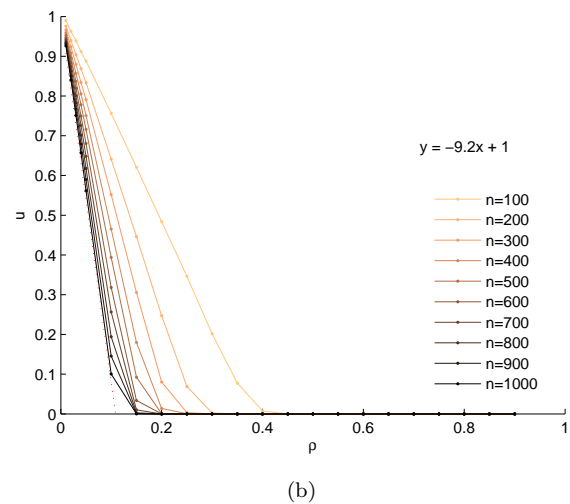}
    \label{fig:fig_rho_u}
}
\caption{(Color online) 
Averages of 20 simulations with various $n$ values where $\nu=10^6$ and $m$ changes as $\rho$ does.
(a) Change of the minimum fame, 
(b) change of the percentage of the completely forgotten agents as $\rho$ changes.}
\label{fig:fig_rho_fmin_u} 
\end{figure}

\subsection{Minimum Fame}
The minimum fame $\bm{f_{min}}$ in the population for a range of parametric settings is 
investigated. As $\rho$ decreases, the minimum value of fame decreases as in 
Fig.~\ref{fig:fig_rho_fmin_u}(a). This decrease turns out to be linear. As $n$ increases the linear 
region becomes more visible and $n$ values $800, \cdots, 1\,000$ produce almost the same line. For 
$n=1\,000$, the line is given as $f_{min}\approx 1.1 \rho -0.12$. The minimum value of fame is $f = 
0$ when the first agent becomes completely forgotten.  Occurrence of the first $f=0$ case depends 
on $n$ and it has quite a dynamic range. The first $f = 0$ case occurs when $\rho$ is around 0.35 
for $n=100$. As $n$ increases, the first $f=0$ case moves to smaller values of $\rho$. For 
$n=1\,000$, it happens at $\rho=0.1$.

%

\subsection{Percentage of Forgotten Agents}
As $\rho$ is decreased beyond the point where at least one agent is forgotten, the minimum fame 
does not provide any further information. For those values of $\rho$, the  number of completely 
forgotten agents can be investigated.

The percentage $\bm{u}$ of the population that is completely forgotten is used and 
Fig.~\ref{fig:fig_rho_fmin_u}(b) is obtained. 
Note that the graphs in 
Fig.~\ref{fig:fig_rho_fmin_u}(a) 
and 
Fig.~\ref{fig:fig_rho_fmin_u}(b) 
complement each other for any particular value of $n$. 
As expected, for any values of $\rho$, 
one graph has non-zero values whenever the other graph has zeros. Here again, as $n$ increases, the 
curves converge to a line which is given as $u\approx -9.2 \rho + 1$ calculated for $n=1\,000$.

\begin{figure}[htbp]
\centering
\subfigure[] 
{
    \includegraphics[width=\columnwidth,height=\columnwidth,keepaspectratio=true]{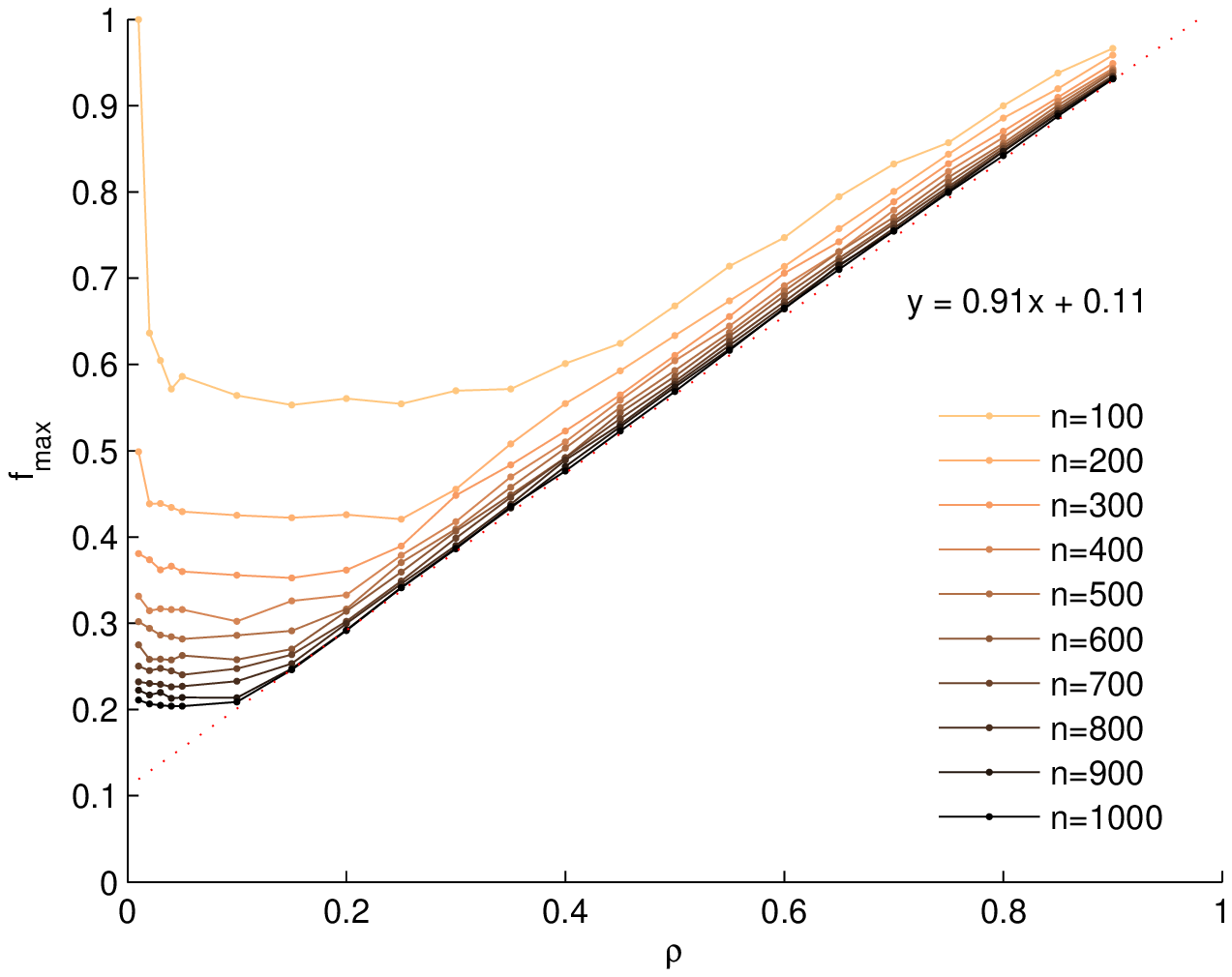}
    \label{fig:fig_rho_fmax}
}
\vspace{.1in}
\subfigure[] 
{
    \includegraphics[width=\columnwidth,height=\columnwidth,keepaspectratio=true]{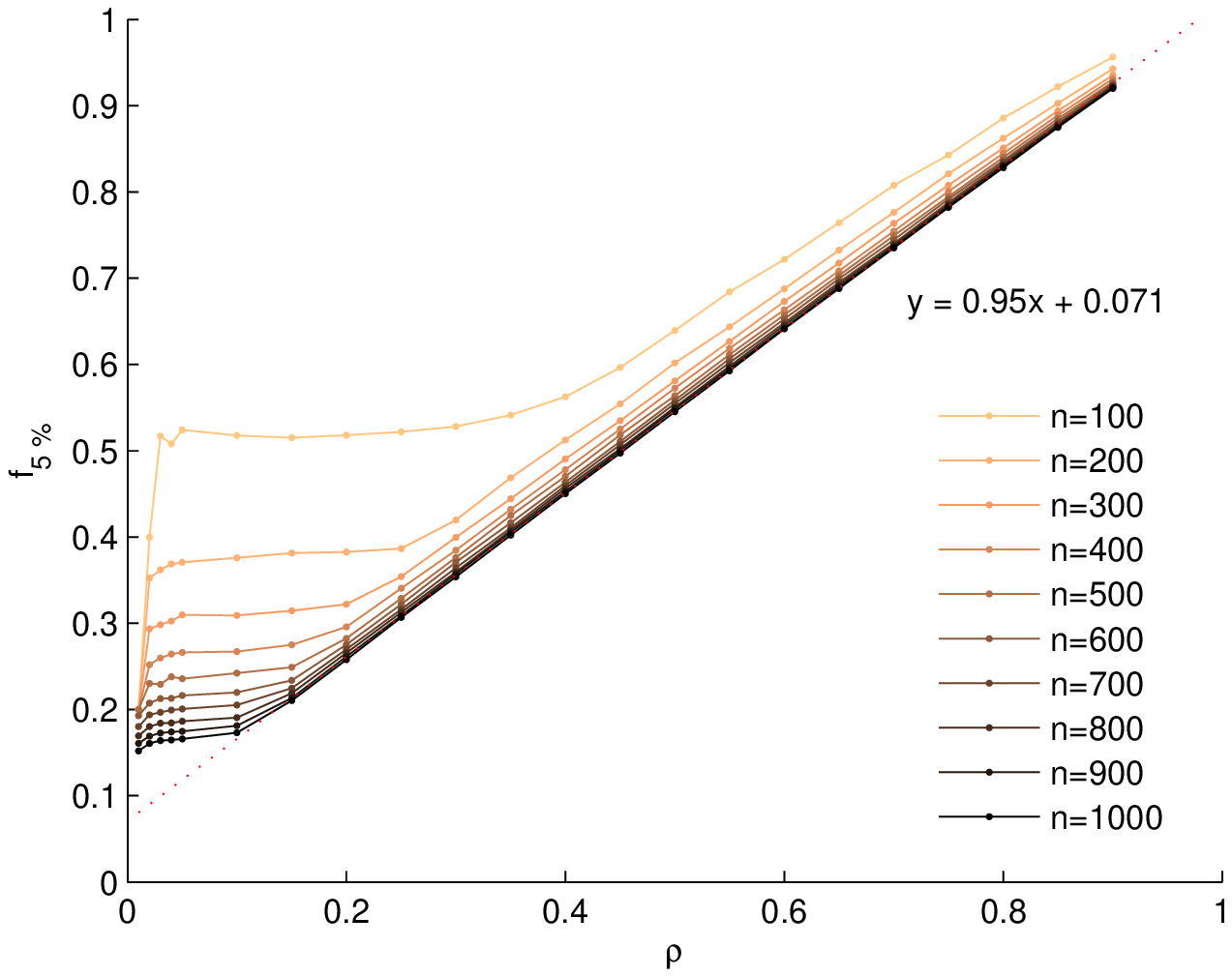}
    \label{fig:fig_rho_ftop5}
}
\caption{(Color online) 
Averages of 20 simulations with various $n$ values where $\nu=10^6$ and $m$ changes as $\rho$ does.
(a) Change of the maximum fame,
(b) change of the cumulative fame of the top 5 \% as $\rho$ changes.}
\label{fig:fig_rho_fmax_ftop5} 
\end{figure}

\subsection{Maximum Fame}
The maximum value of fame $\bm{f_{max}}$ has an interesting behavior as $\rho$ changes. For 
$n=100$, $f_{max}$ slowly decreases as $\rho$ decreases from 0.9 to 0.1. It reaches a minimum value 
around $\rho = 0.1$. Interestingly, further decrease of $\rho$ causes  $f_{max}$ to increase. This 
pattern can be seen by tracing 
Fig.~\ref{fig:fig_rho_fmax_ftop5}(a) 
from right to left, where the 
emergence of fame can be observed as the relative memory size (indicated by $\rho$) is decreased. 
This unexpected behavior can be explained: When $\rho = 1$, every agent is known by everybody else 
so the fame is 1. As $\rho$ decreases, the memory size of the agents decreases. Since no one 
dominates the memories yet, people are almost evenly distributed in the memories. So the reduction 
of the maximum fame is due to the decrease of the memory size. But as $\rho$ keeps decreasing, 
after a certain point some people become completely forgotten and some others become the dominating 
ones. As $\rho$ approaches to the limit of 0, more people become completely forgotten and fewer 
people dominate the memories. Those that dominate take all the references. So the rapid increase of 
maximum fame in the vicinity of $\rho=0$ can be explained due to this positive feedback.

Another observation is that the decrease of maximum fame is also linear. As in the case of minimum 
fame, as $n$ gets larger, the linear pattern becomes more apparent. For $n=1\,000$, it is given as 
$f_{max} \approx 0.91 \rho + 0.11$.


\subsection{Cumulative Fame}
Maximum fame is a measure of the dominance of one agent. Dominance of a group of famous agents is 
investigated by means of the cumulative fame of the top $\bm{p}$ percent of agents ordered 
according to their fames. The top $p\%$ of the population is selected. The \textit{cumulative fame} 
$\bm{f_{p\%}}$ is obtained by adding their fames. The maximum possible value for the cumulative 
fame is $np/100$ when all top $p\%$ are completely known, that is, each has fame of $f=1$. This 
value is used for normalization. 

In this study, $p=5$ is used and 
Fig.~\ref{fig:fig_rho_fmax_ftop5}(b) 
is obtained. 
As $n$ increases, the curves converge to a line which is given as $f_{5 \% } \approx 
0.95 \rho + 0.071$ for $n=1\,000$. 
Fig.~\ref{fig:fig_rho_fmax_ftop5}(a) and
Fig.~\ref{fig:fig_rho_fmax_ftop5}(b) are
quite similar as expected. 
On the other hand, $f_{5 \% }$ line decreases slightly faster than 
$f_{max}$ line as $\rho$ decreases. As $\rho$ decreases to $\rho=0.1$, the curves become saturated. 
They stay this way for awhile and then as $\rho$ approaches 0, they start to decrease again. This 
behavior near $\rho=0$ can be explained by the memory size. As $m$ decreases, at some point there 
is no space to keep 5\,\% of the population. Whenever that happens, the cumulative fame of the top 
5\,\% starts to decrease towards 0. This final decrease is much sharper. This behavior can be seen 
more clearly for small values of $n$ in the figure. For example for $n=100$, top 5\,\%, means 5 
agents. If $m$ becomes less than 5, that is $\rho<0.05$, a sharp decrease is expected as in the 
figure.


\section{Network Issues}

Who-knows-who graph is another representation of the state of the system. The directed graph 
$G(A,E)$ where $A$ is the set of agents and $E = \{ (a_i, a_j) \: | \: a_j \in M_i \}$ is called 
the \textit{who-knows-who graph}. The graph is a directed graph, since the corresponding relation 
``to know'' is not symmetric.

In this directed graph, out-degree is not interesting since all vertices have the same out-degree 
of $m$, independent of recommendations. On the other hand, in-degree of a vertex changes by the 
recommendations and has a dynamic range starting from 0 and it can be as large as $n$. For both 
random and regular initial memory cases, the initial in-degree distribution is uniform since every 
agent has the same knownness of $m$. As a result of recommendations, in-degrees of a few agents 
increase while the majority decreases to 0. So as a result of recommendations, uniform in-degree 
distribution degrades to the one with two peaks around 0 and $n$. At an absorbing state, there 
would be exactly two non-zero points in the in-degree distribution, namely 0 and $n$. There is a 
nucleus of $m$ vertices in which a vertex is connected to other $m-1$ vertices and itself. The 
remaining $n-m$ vertices are connected to this $m$-vertex nucleus. The $m$ vertices of the nucleus 
have in-degree of $n$ and $n-m$ vertices have 0.

The undirected graph underlined by the directed who-knows-who graph is topologically investigated. 
In the random case, the initial network is a random graph. In the regular case, the initial graph 
is regular. As the recommendation dynamics takes place and fame emerges, both initial graphs 
transform into one common graph structure. The few famous vertices, which are in the process of 
forming the nucleus, become hubs. The rest of the vertices are connected to these hubs. Giving this 
picture, the graph is more towards to star-connected rather than power-law degree distribution. 
Therefore, the average distance is very low. The clustering coefficient is also low. The 
recommendation is a transformation that uses local information only. There are some network growth 
models that also use local information only but they produce power-law degree 
distribution~\cite{Krapivsky2001, Rozenfeld2004}.

\section{At the absorbing states} 
The main focus of this work is the behavior of the system as it approaches but never reaches to an absorbing state.
In this section a brief investigation of the system at the absorbing state is done and the rest is left as future work.
The system simply takes too long time to reach an absorbing state for $n$ values considered so far. 
On the other hand, if one reduces $n$, 
absorbing states become obtainable within reasonable durations. 
Simulations are done for small values of $n$ and $m$ such as $n \in \{ 20, 30, 40, 50, 60\}$ 
and $m \in \{ 1, 2 \}$.
As a measure of time, the number of simulation cycles required to reach an absorbing state is measured.

It is clear that the system reaches an absorbing state asymptotically. 
Therefore, near absorbing state, forgetting the next person becomes increasingly harder.
Let $t_{i}$ be the time the $i$th person is forgotten.
Then, define the \textit{time needed to forget the next person} after the $i$th person as  $\Delta t(i) = t_{i+1}-t_{i}$ where $i \in \{ 1, \cdots, n-m\}$.
As expected, $\Delta t(i)$ rapidly increases as $i$ approaches to $n-m$. 

\begin{figure}
\includegraphics[width=\columnwidth,height=\columnwidth,keepaspectratio=true]{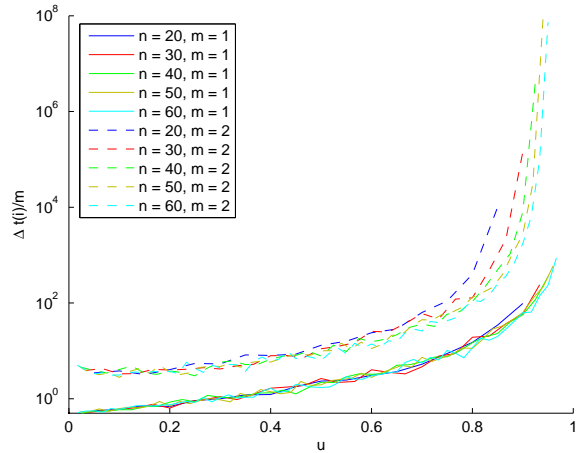} 
\caption{(Color online) Time to forget next person scales with $n$ for small $n$ and $m$ values. Average of 40 simulations.}
\label{fig:fig_absorbingStateA1B1}
\end{figure}

In some systems, system size makes a big effect to the behavior.
When the parameters are scaled with the systems sizes, then some regularities become visible~\cite{Hinrichsen2000}.
In this system $n$ turns out to be an important parameter.
Fig.~\ref{fig:fig_absorbingStateA1B1} provides the behavior of $\Delta t(i)$ as average of 40 simulation runs.
In the X-axis the number of persons forgotten is scaled by $n$ which is the percentage of forgotten, that is $u$.
In the Y-axis $\Delta t(i)$ is scaled as $\Delta t(i)/m$ for various values of $n$.

Interestingly, there are two family of curves.
The upper family belongs to $m=2$. 
It starts with a slight decrease which corresponds to the initial trend of forgetting.
Then the regime changes and forgetting becomes harder and harder as $u$ approaches to 1.
The $m=1$ family does not have this pattern.
Unfortunately, these $n$ and $m$ values are too small to observe the patterns that are focused in this work.

\section{Related Work on Fame} 
Recently some studies on fame has been done. The difficulty starts with the definition of fame. An 
innovative metric for fame is defined as the number of hits returned from a search of a person's 
name on Google~\cite{Simkin2006a, Simkin2006b}. In this study, the fames and the achievements of 
WWI fighter pilots are examined. ``Fame'' $F$ is defined as the number of Google hits. 
``Achievement'' is the number of opponent aircrafts destroyed. It is found that fame grows 
exponentially with achievement. The distribution of fame is given as $P(F)\propto F^{- \gamma}$ 
where $\gamma\approx 2$. A study on scientist using the number of Google hits for fame gives 
another distribution, $P(F)\propto e^{-\eta F}$ where $\eta=0.00102$~\cite{Bagrow2004, Bagrow2005}.

Scientific papers can be ``famous'' by getting cited. A study on scientific papers published in 
Physical Review D in 1975-1994 has been done~\cite{Simkin2003a}. There are 24\,000 papers, 350\,000 
citations, that is 15 citation per paper on the average. Yet, 44 papers are cited 500 times or 
more.  It is found that copying from the list of references used in other papers has an impact. A 
paper that is already cited has more chance to get cited again.

Early results of the simple recommendation model such as the fast and slow forget regimes, 
asymptotic approach to absorbing states and degeneration of the distribution of knownness to fame 
as $\rho$ decreases were presented in~\cite{Bingol2005}.

\section{Conclusions}
``Too many to remember'' is quite the common case in many complex systems. A dynamic memory model 
is defined where agents interact by exchanging recommendations. A random-selection based model is 
described as the simplest instantiation of the general model. Although the model does not prefer 
any agent, some agents become increasingly famous as the memory gets smaller. This observation can 
be interpreted as the \textit{emergence of fame}.

The model can be used in some practical applications. Suppose some agents are preferred in the 
recommendations. Then, their fame is expected to increase and last longer. This can be used to 
model the social dynamics of advertisement. Essential questions in advertisement such as how 
frequently advertise or how widely advertise can be better estimated. Voting or election results 
are studied in opinion dynamics~\cite{SznajdWeron2000}. Emergence of fame can be considered as 
formation of an opinion through interactions of agents.

The general model will serve as a basis for building sophisticated models as different selection 
criteria are adopted and the agent interaction scheme is restricted with new assumptions. For 
example, it is possible to define selections so that only agents with a common friend can interact. 
This leads to a version of Axelrod's Culture Model~\cite{Axelrod1997}. The model can be modified so 
that the giver always recommends itself rather than some agent from its memory. Then it becomes 
very close to the small-world model presented in \cite{Watts1998}. Another possibility is to place 
the agents on the vertices of an interaction graph, possibly with small-word or scale-free 
properties in order to introduce real world flavor.

\acknowledgments 
This work was partially supported by 
Bogazici University Research Projects under the grant number 07A105
and was partially based on the work performed in the framework of the FP6 project SEE-GRID-2, 
which is funded by the European Community (under contract number INFSO-RI-031775).
The author would like to thank Sidney Redner, Alber Ali Salah, Cem Ersoy, Can Ozturan and Pinar Yolum 
for their useful comments. 
The author also thanks to Muhittin Mungan, Amac Herdagdelen for their invaluable support especially in scaling
and to anonymous referees for their enriching comments and suggestions.

\bibliography{bingolFame}

\end{document}